\documentclass[12pt,a4paper]{article}

\usepackage{amsfonts} 
\usepackage{amssymb}
\usepackage[dvips]{graphicx}
\usepackage{epsfig}
\newcommand{\beq}{\begin{equation}}
\newcommand{\eeq}{\end{equation}}
\newcommand{\p}{\partial}

\begin{document}

\hfill SPIN-04/29

\hfill ITP-UU-04/17

\vspace{30pt}

\begin{center}
{\Large \bf Polygon model from first order gravity}

\vspace{20pt}

{\sl Z. K\'ad\'ar }\\

\vspace{24pt}

{\footnotesize
Institute for Theoretical Physics, Utrecht University,\\
Leuvenlaan 4, NL-3584 CE Utrecht, The Netherlands\\ 
{\tt email:} {\tt z.kadar@phys.uu.nl}\\
}
\end{center}

\vspace{1pt}

\begin{center} 
{\bf Abstract}
\end{center}

The gauge fixed polygon model of 2+1 gravity with zero cosmological constant and arbitrary number of spinless point particles is reconstructed from the first order formalism of the theory in terms of the triad and the spin connection. The induced symplectic structure is calculated and shown to agree with the canonical one in terms of the variables.
\section{Introduction}
In the beginning of the 90's the gauge-fixed polygon model of $2+1$ dimensional gravity with zero cosmological constant constructed by 't Hooft appeared in the literature \cite{causality}. The motivation was to show that closed timelike curves cannot appear in a universe with the topology $M^3=M^2\times\mathbb{R}$, started off with good initial conditions and containing only spinless point particles. Soon it was discovered that the model gives rise to an interesting dynamical system with surprising complexity near the Big Crunch and that the direct geometric meaning of the phase space variables may enable one to make predictions about the possible appearance of the initial and/or final singularity \cite{evolution,gth}. There, the studies were based on numerous computer simulations, where the universes were constructed by intelligent guesses of the not independent phase space variables such that all constraints among them admit solutions. Later, it has been noticed \cite{canquant}, that if the Hamiltonian is taken to be the total curvature of the spacelike slice expressed in terms of the phase space variables, then it generates the correct time evolution if the symplectic structure is the canonical one ($p_idq_i$). This was a promising discovery with respect to quantization, and it lead to a qualitative analysis of the possible spectrum of the variables in the quantum theory. However, due to the nonpolynomial form of the Hamiltonian and appearance of the discrete gauge artifacts of the model called "transitions" during the otherwise trivial time evolution of the phase space variables, the quantization was not completed.

There was more progress in quantization of pure gravity, because in \cite{achtow,witt} it has been discovered that $2+1$ gravity - apart from important subtleties \cite{matschullehhp} - is a gauge theory of a connection, which is composed of the triad and the spin connection. The extension to the inclusion of particles was readily made \cite{soge}, but the research mainly concentrated on the case of pure gravity (see \cite{carlip} for an overview), apart from a few papers where path integral \cite{menotti}, combinatorial quantization \cite{bernd} or lattice methods \cite{waelmod} have been applied for treating particle degrees of freedom coupled to gravity. The smooth first order formalism was studied extensively by Unruh and Newbury\footnote{Thanks to P. Menotti for this reference.} \cite{unne} (and quantized if one particle is present) by Matschull and Welling \cite{matschullonepart,matschullmp}, see also \cite{erickarim}. The idea in these approaches is a feature of three dimensions that one can reduce the infinite dimensional space of the physical fields to the space of a finite number of variables.  Note also, that with a few exceptions \cite{exceptions,unne}, explicit results mainly correspond to topologies where $M^2$ has genus less than two or noncompact, even though topology change is not a priori forbidden in a quantum theory \cite{witttc}.  
The different approaches are usually difficult to compare and often they provide different quantizations. In particular, the relation of the 't Hooft polygon model to other approaches is poorly understood. It is a representation of Poincare group ($ISO(2,1)$) structures on Minkowski space so it is naturally a second order approach to $2+1$ dimensional General Relativity. As opposed to the mentioned first order approaches \cite{unne,matschullmp}, it contains an explicit time slicing in terms of piecewise flat Cauchy surfaces with some conical singularities which do not necessarily correspond to point particles (which are {\em 3d} curvature singularities). This is different both from the York gauge, where the slices are characterized by the value of the trace of the extrinsic curvature and the conformal gauge where the extrinsic curvature of the slices vanishes. The arising dynamical system has an explicitly written Hamiltonian for arbitrary genus of $M^2$ and arbitrary number of non static point particles.
Recently, there has been progress in the spin foam quantization and loop quantum gravity, where spinning particles were also included \cite{freidel,karimperezpart}. The results there are related to the polygon model; it can also be derived from the first order variables, and the the fundamental object describing space time is a decorated graph similarly to the spin networks of the mentioned loop and foam approaches. 

The comparison with a triangular lattice version of the first order formalism has been done by Waelbroeck and Zapata \cite{thooftwael}. 
It is worked out for the smooth case in this article. The first step is a choice of graph $\Gamma$ in space $M^2$ such that the the complement is contractible. Inside, the flat connection is given by a pure gauge, and the system reduces to the dynamics of the "edge" vectors and the Lorentz holonomies, specifying the contractible region in $M^2$ and the identifications at the boundaries. The second step is a gauge transformation to make all edge vectors lie in a spacelike surface. Then one has to show that only the lengths of the vectors and the so called boost parameters of the holonomies are independent, the symplectic structure reduces to the canonical one and the remaining constraint algebra closes, no second class constraints appear.

In the next section an introduction is given to the model with special emphasis on the particle content and the phase space structure. Section 3. is a brief summary of the first order formalism, Section 4. contains the reduction of fundamental fields to a finite number of covariant variables, which were essentially done in \cite{matschullmp} for trivial tangent bundle of the spacelike slice. Finally the last section contains the main result of the paper: the reduction of the covariant variables to those of the 't Hooft polygon model, the induced symplectic structure is recovered both directly from the fields and from the covariant variables and the dynamics is described.                    
\section{Polygon model} \label{poligonmod}
In this section an introduction is given to the model based on the so called {\em one-polygon tessellation}. It applies to space times which admit a slicing in terms of one spacelike Euclidean polygon. In other words the space $M^2$ has a piecewise flat 2-metric with some conical singularities. There exists no rigorous proof that all solutions to the Einstein equations with pointlike spinless sources admit such slicing. However, there is ongoing work establishing this result in the mathematics literature \cite{barbot} for the case of matter free universes, see also \cite{mi}. In the general case one needs more polygons with appropriate pasting conditions. However, the phase space structure of the model can be described in a compact way with one polygon, that is with a graph embedded in $M^2$ such that its complement has one component. The discussion of the general case is technically more difficult, but it contains no new physics. The phase space variables are associated to the edges of this graph and the edges of the dual graph, which are closed curves in this case. 

One may need to consult the original papers \cite{causality,evolution} for throrough physical interpretation and intuition behind the construction. Here the introduction is similar to reference \cite{mi}, where only the matter free case was discussed, although the emphasis here is on the the gauge fixing: the fact that the spacelike slice is a {\em planar} polygon makes all but one scalar parameter redundant of both the edge vector and the Lorentz holonomy. 

Suppose that the $M^3=\Sigma_{g,N}\times I$ is a globally hyperbolic flat Lorentzian manifold with a Cauchy surface of the topology $\Sigma_{g,N}$: a Riemann surface of genus $g$ with $N$ number of punctures corresponding to (spinless) particles and $I$ is an interval. For $g=0$ suppose that $N>2$ and for $g=1$ suppose that $N>0$. In the following, we describe the initial value Cauchy surface and its parametrization, then explain how it arises. 
\subsection{Phase space}
\begin{figure}
\begin{center}
\includegraphics[width=8cm,height=3cm]{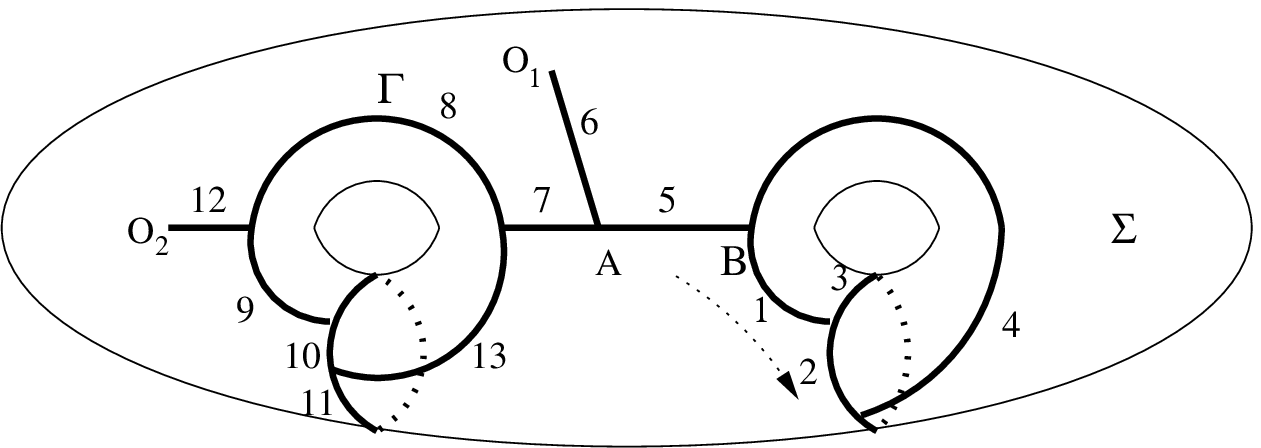}
\quad
\includegraphics[width=3cm,height=3cm]{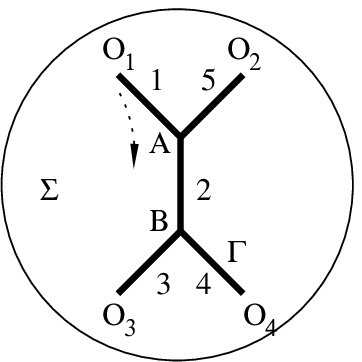}
\vspace{0.5cm}

\includegraphics[width=8cm,height=5cm]{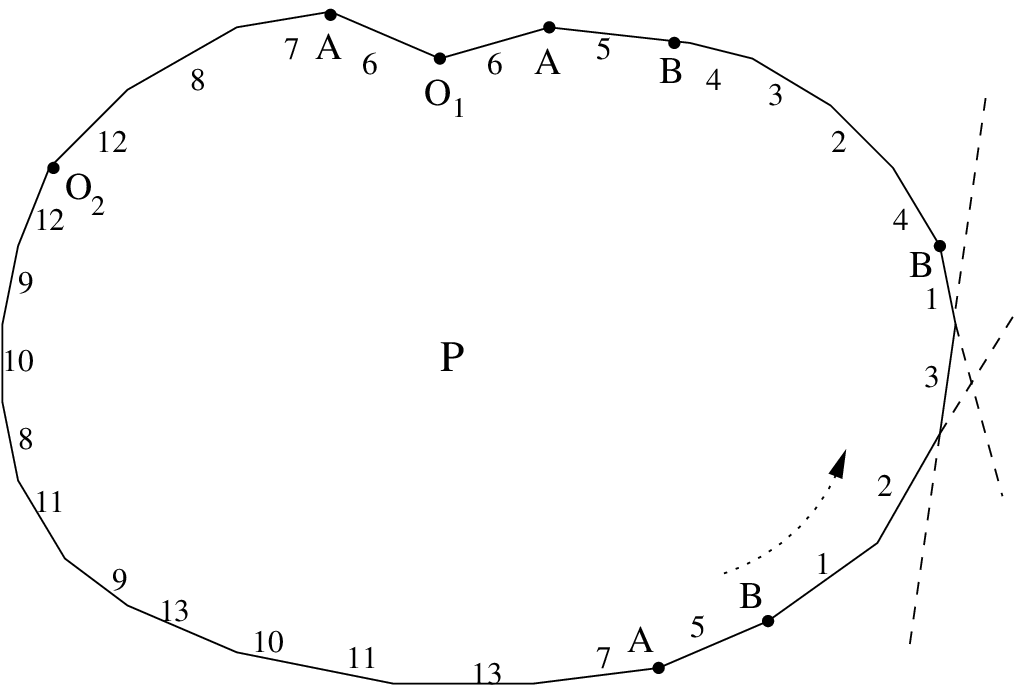}
\quad
\includegraphics[height=5cm]{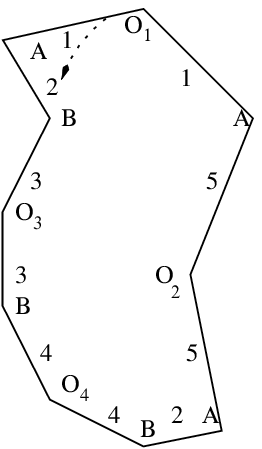}
\vspace{0.5cm}

\includegraphics[width=9cm,height=3.5cm]{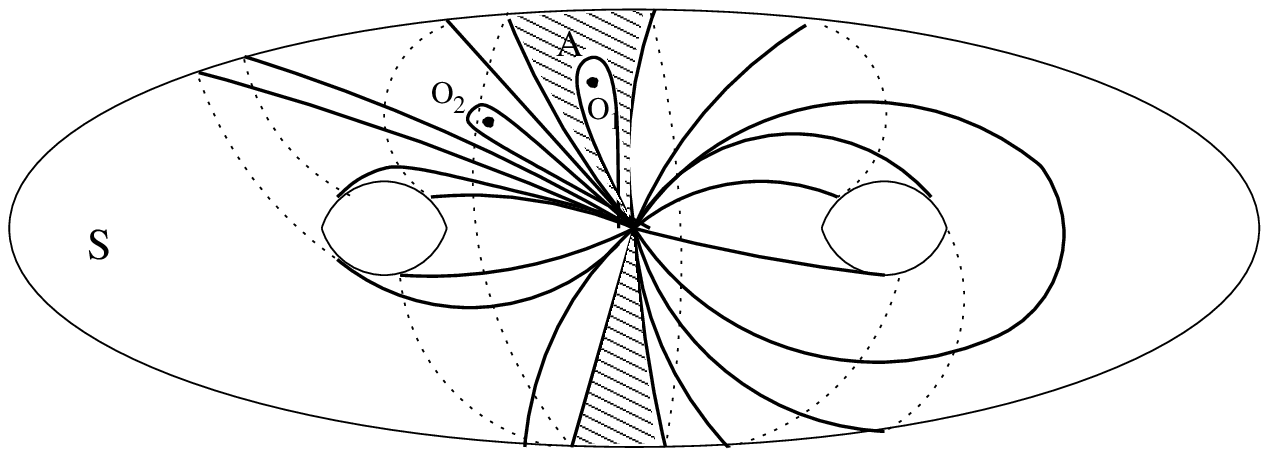}
\quad
\includegraphics[height=3.5cm]{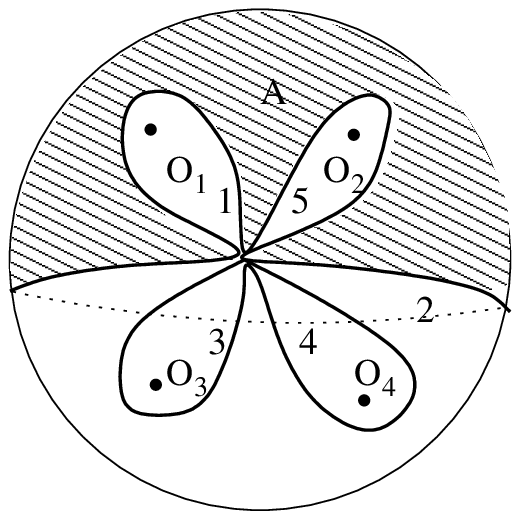}
\caption{\label{onepol}{Examples of a $g=2$, $N=2$ and $g=0$, $N=4$ surfaces $\Sigma$ with the fat graph $\Gamma$ above, the corresponding polygons $P=\Sigma\backslash\Gamma$ in the middle and the dual graphs in the bottom (Note that the dual graph of the genus two surface on the bottom left is reflected with respect to the plane of the paper). The few illustrated dashed lines are the intersections of the hyperplanes given by (\ref{wall}) with the $t=0$ spacelike planes. The edges of the polygon $P$ are determined by the segments between the intersections of these lines.
Those edges of the polygon bearing the same number have the same length and are to be glued with opposite orientation. The points $A$ and $B$ are examples of 3-vertices, $O_i$ are 1-vertices. At vertices the sum of the angles are different from  $2\pi$. In the bottom figures the triangle bounded by the dual edges corresponding to those edges of $\Gamma$ which are incident to the 3-vertex $A$ are shaded. The dual graph cuts the surface into triangles and discs, corresponding to 3-, and 1-vertices, respectively.}}
\end{center}
\end{figure}

The phase space variables of the model are encoded in a decorated fat graph $\Gamma$ on the surface $\Sigma_{g,N}$ (to be called $\Sigma$ from now on). The graph satisfies two properties. 
\begin{itemize} 
\item The complement $P=\Sigma\backslash\Gamma$ is simply connected and has only one component.
\item It is a trivalent graph except for $N$ edges ending at the punctures.
\end{itemize}
In other words $\Gamma$ has one face $F=1$, then consequently the number of its edges is $E=6g-3+2N$ and the number of its vertices is $V=4g-2+2N$. We denote the vertices to which three edges are incident by {\em 3-vertices} and those ending at punctures by {\em 1-vertices}. Then for $N=0$ one has $2E=3V$ and $F+V-E=2-2g$. The induced metric on $P$ is Euclidean, it is a geodesic polygon. The initial value surface is $P$ modulo gluing prescribed by the (cut) fat graph which is the boundary: $\p P=\Gamma$. Examples are given in Fig.\ref{onepol}. 
In other words the surface $\Sigma$ is endowed with a locally flat metric, with conical singularities at the vertices of $\Gamma$. The decoration of $\Gamma$ is an assignment of one real and one positive number $\eta$ and $L$, respectively, to each edge which characterizes the Euclidean structure completely in the following way:
The numbers $L$ denote the lengths of the edges. The numbers $\eta$ encode the angles of the polygon in such a way that any three angles corresponding to a 3-vertex of $\Gamma$ are (essentially) the angles of a hyperbolic triangle with edges $2\eta_i,2\eta_j,2\eta_k$, where $i,j,k$ denote the edges incident to the vertex. The angles at 1-vertices of $\Gamma$ are determined by the mass of a corresponding particle and the parameter $\eta$ associated to the corresponding edge. The precise relations will be specified below. We will need the dual graph $\gamma$ of $\Gamma$. It has one vertex $p$ inside $P$ and each edge $\zeta_i$ is the simple closed curve on $\Sigma$ from $p$ to $p$ such that it intersects edge $i$ of $\Gamma$ and it intersects only that edge. We choose an orientation for all dual edges of $\gamma$ and extend the decoration of $\Gamma$ by denoting the side of edge $i$ where the dual edge is oriented toward the other side by $i_+$ and the other side by $i_-$. That is, all edges of the $12g-6+4N$ sided polygon $P$ has a label and those with $i_+$ and $i_-$ are identified with opposite orientation. The quotient is the manifold $\Sigma$. 
Remember, that a 1-vertex is a puncture, that is the dual edges corresponding to the edges of $\Gamma$ ending at one vertices are also noncontractible. Then the homotopy class (with endpoints fixed) of a dual edge is unique, no two distinct dual edges are in the same homotopy class and no dual edge is in the trivial class. $\Sigma\backslash\gamma$ consists of $N$ discs and $4g-2+N$ triangles.        

Where does the above structure come from? A solution to the Einstein equations is a locally flat manifold with conical curvature singularities at the world lines of the particles, such that these are timelike (lightlike) geodesics if the particle is massive (massless). It is possible to specify such a manifold as a geometric structure: $M$ is locally modeled on $\tilde{M}={\mathbb R}^{2,1}$ with transition functions in the isometry group $ISO(2,1)$ of $\tilde{M}$ (${\mathbb R}^{2,1}$ denotes Minkowski space, ISO(2,1) is the Poincare group). The {\em holonomy map}: 
\beq \rho:\pi_1( M)\to ISO(2,1) \eeq
is associated to the construction: it is a homomorphism. We are considering globally hyperbolic manifolds with topologies $M=\Sigma\times I$, for which $\pi_1(M)=\pi_1(\Sigma)$, so the homotopy classes in $M$ are identified with the homotopy classes in $\Sigma$.  
The holonomy map assigns Poincare group elements to each edge of $\gamma$. Since the number of its edges is more than the number of independent loops in a canonical homology basis of the fundamental group ($E=6g-3+2N>2g+N-1$), the set of generators is overcomplete, there are relations among them. 

Having the last paragraph said, we are ready to describe the explicit construction. Take a point $Q$ in Minkowski space given by the origin of the Cartesian coordinate system $X=(t,x,y)$. (It is a preimage of a point in $M$ under the universal cover.)
One of its images ${\cal P}_i(Q)\equiv \rho(\zeta_i)(Q)\equiv Q_i$ is given by
the origin of another Lorentz frame $X'=(t_i,x_i,y_i)$ such that $X'={\cal P}_i X$. (Let us remind the reader 
that the dual edge $\zeta_i$ is a closed curve, that is an element of the fundamental group.) Define hyperplanes in $X$ by
\beq t=t_i\equiv\left({\cal P}_{i\pm}X\right)^0\equiv\left({\cal P}^{\pm 1}_i X\right)^0. \label{wall} \eeq
These have spacelike normal vector. The intersection of these hyperplanes with the $t=0$ plane in $X$ are straight lines. Choose a circular ordering of all sides of the edges of the fat graph $\Gamma$. The straight lines bound a polygon the edges of which are the intersections of the previously defined straight lines with their neigbours according to the circular ordering. For clarity let us summarize the procedure again. To each side of each edge of $\Gamma$ there is a corresponding oriented dual edge, to each oriented dual edge there is a corresponding Poincare holonomy, each Poincare holonomy maps point $Q$ to point $Q_i$ and equations (\ref{wall}) and $t=0$ define a straight line. These straight lines bound a polygon\footnote{Proving that starting from any geometric structure describing a globally
hyperbolic Lorentzian spacetime with the given conical singularities one can construct a polygon without bad self intersections \cite{evolution} would amount to proving the statement that any universe with $N\neq 0$ admits one polygon tessellation. We do not have such a proof.} $P$ which contains $Q$. See also edge $3$ in the middle left part of Fig.\ref{onepol} for illustration. 
One can do the same procedure starting from $Q_i$ and obtain an isometric polygon meeting the previous one at edge $i$ and having a timelike normal vector different from the purely timelike $n=(1,0,0)$ of $P$ containing $Q$. In fact each element of the fundamental group gives another copy of the polygon. This way we obtain a universal branched covering of the surface $\Sigma$ with branch points at the vertices. For the case $N=0$ it is useful to visualize the situation as a nonplanar tiling of $\mathbb{R}^2$ with the polygon $P$ such that two copies meet at edges and three copies meet at vertices and the orientation is opposite for any two adjacent copies. The length $L_{i\pm}$ is the length of the edge given by the purely spacelike three-vector $E_{i\pm}$ as the difference of the two vertices bounding the edge $i\pm$ in coordinate system $X$. $L_{i+}=L_{i-}\equiv L_i$ by construction. The angles of the polygon can be determined from the consistency condition we have for each vertex of $\Gamma$. Assume that edge $j,k,l$ are incident to a 3-vertex. Then
\beq {\cal P}_j{\cal P}_k{\cal P}_l=id, \eeq
since for the dual edges $\zeta_j\circ \zeta_k\circ \zeta_l=e$, where $\circ$ denotes multiplication in the fundamental group and $e$ is the unit element. Of course $j$ above is either $j_+$ or $j_-$ (and $k,l$ likewise), one has to be 
careful about the orientation. After decomposition of the Poincare transformation as 
\beq X_i={\cal P}_iX=\Lambda_i X+a_i \label{poincare} \eeq
with $\Lambda_i \in SO_+(2,1)$ ($+$ denotes the identity component, the restriction means allowing only orientation preserving Lorentz transformations), and $a_i \in {\mathbb R}^3$, we get
\beq \Lambda_j\Lambda_k\Lambda_l=id \label{consist} \eeq
and another relation, which involves also the $a_i$'s. Using the definition of the boost and the rotation:
\beq
B(\xi)=\left( \begin{array}{ccc}
\cosh \xi & 0 & \sinh \xi \\
0 & 1 & 0 \\
\sinh \xi & 0 &  \cosh \xi
\end{array} \right), \quad
  R(\psi)=\left( \begin{array}{ccc}
1 & 0 & 0 \\
0 & \cos \psi & -\sin \psi \\
0 & \sin \psi & \cos \psi \end{array} \right).
\label{rmatrix}
\eeq
we can parametrize the Lorentz group elements as 
\beq \Lambda_i=R(\phi_i)B(2\eta_i)R(\phi'_i) \label{param}  \eeq
and the three angles at the vertex are given by
\beq \alpha_j=\phi'_k+\phi_l\quad\alpha_k=\phi'_l+\phi_j\quad\alpha_l=\phi'_j+\phi_k. \eeq
since the formula (\ref{consist}) rewritten with the help of (\ref{param}) means that the parallel transport around a vertex can be written as a sequence of rotations and boosts. Note that the angle parameters $\phi$ of the Lorentz group elements can be chosen without loss of generality in such a way that the angles $\alpha \in [0,2\pi)$ at the vertices. The geometric picture is the following. The polygon containing $Q$ meets the polygon containing $Q_l$ and that containing $Q_{k\circ l}$ at the vertex. If the normal vector of the first is denoted by $n$ then $\Lambda_l n$ is that of the second and $\Lambda_k\Lambda_l n=\Lambda^{-1}_j n$ is that of the third. The hyperbolic ``angles" between the polygons at the edges are the boost parameters $2\eta_j\,,2\eta_k\,,2\eta_l$. The relation $n=\Lambda_j \Lambda_k \Lambda_l n$ says that the parallel transport around the vertex does not change the vector $n$: space time is flat at the vertex. The group $SO_+(2,1)$ is the (orientation preserving) isometry group of the hyperbolic space ${\mathbb H}_2=\{(t,x,y):-t^2+x^2+y^2=-1\}$ with the induced metric from Minkowski space with constant negative curvature. The normal vectors are points in that space and the relation (\ref{consist}) give the angles $\pi-\alpha_j\,,\pi-\alpha_k\,.\pi-\alpha_l$ of a triangle\footnote{The precise identifications of angles of a hyperbolic triangle in terms of the angles for the case when, say, $\alpha_j>\pi$ is explained in \cite{mi}. There can be at most one concave angle at a vertex \cite{gth}, which makes the functions $\alpha(\eta_j,\eta_k,\eta_l)$ unambiguous for all three incident angles.} in terms of its lengths $2\eta_j\,,2\eta_k\,,2\eta_l$. We still need to specify the angles at 1-vertices. The normal $n$ is mapped to $\Lambda_i n\neq n$ under the holonomy, thus a 1-vertex carries nonzero (three dimensional) curvature. The angle there is 
\beq \phi_i+\phi'_i\in [0,2\pi), \eeq
since the parallel transport around a 1-vertex is a boost and a rotation with the above angle. 
\subsection{Constraints, degrees of freedom, symmetries}
In terms of the variables $\Lambda_i$ and $E_i$  we have constraints. 
\begin{itemize}
\item Half of the edge vectors are not independent since  
\beq E_{i+}=-\Lambda_i E_{i-}, \eeq
where the minus sign is due to the opposite orientation. Using the decomposition of (\ref{param}), one can write explicitly:
\beq E_{i-}=\pm L_i R(-\phi'_i)e^{(1)} \quad E_{i+}=\mp L_i R(\phi_i)e^{(1)} \label{edgevec} \eeq
with $e^{(1)}=(0,1,0)$. 
\item There are $4g-2+N$ consistency conditions of the type of eqn.(\ref{consist}) corresponding to 3-vertices. 
\item There is a global constraint corresponding to the closure of the polygon $P$:
\beq \sum_i(E_{i+}+E_{i-})=0. \label{closure} \eeq
\item There are $N$ consistency constraints corresponding to 1-vertices.
\end{itemize} 
For the description of the latter we need the definition of the 
{\em axis} of a Lorentz transformation $\Lambda$. The axis is the eigenvector of $\Lambda$ corresponding to eigenvalue $1$ and it reads:
\beq p=c(\cosh\eta \sin\frac{\phi+\phi'}{2},\; -\sinh\eta \cos\frac{\phi-\phi'}{2},\; -\sinh\eta \sin\frac{\phi-\phi'}{2}), \label{axis} \eeq
and for later convenience we fix the constant $c$ to be 
\beq c=sgn\sin\frac{\phi+\phi'}{2} \eeq
whenever the rhs. is nonvanishing.
If it is a timelike (lightlike) vector, the transformation is called elliptic (parabolic). If it is elliptic, then it is conjugate to a pure rotation 
\beq \Lambda_i=gR(2\pi-m)g^{-1} \label{elliptic} \eeq 
with $g\in SO_+(2,1)$. The parameter $m$ is interpreted as the mass\footnote{The definition of the mass comes from writing $T^{00}=m\delta(x-x(t))$ for the static particle in the Einstein equations \cite{djh}.} of the particle in units $8\pi G=1$, where $G$ is Newton's constant. Taking the trace of (\ref{elliptic}), after elementary manipulations, we find:
\beq \cos\frac{m}{2}=\pm\cosh\eta\cos\frac{\alpha_{def}}{2} \label{massshell} \eeq
where $\alpha_{def}=2\pi-(\phi+\phi')\in(0,\pi)\cup(\pi,2\pi)$ is the deficit angle at a 1-vertex with a corresponding elliptic $\Lambda_i$. The equation (\ref{massshell}) can be rewritten in the following way:
\beq p^2+\sin^2\frac{m}{2}=0, \label{massshell2} \eeq
where the square means the Minkowski scalar product with signature $(-++)$.
The above formula is the analog of the mass shell condition for the particle. To see that $p$ is indeed the momentum of the particle, note that the spacelike direction of the velocity vector of the 1-vertex is given by 
\beq E_{i-}-E_{i+}\sim \left(0,\cos\frac{\phi_i-\phi'_i}{2}, \sin\frac{\phi_i-\phi'_i}{2}\right)\sim (0,p^1,p^2). \eeq
Taking into account eqn.(\ref{massshell}) we can conclude that the axis $p$ is indeed the momentum of the particle\footnote{The unusual sine in the mass shell equation is a matter of convention \cite{matschullonepart}. It reflects the fact that 2+1 gravity is a toy model and the mass parameter coming from the stress energy tensor will appear with a unusual sine function in the mass shell condition. This analysis also shows, that arbitrary negative masses may be permitted since all the particles follow timelike geodesics.}, the sign above gets fixed by choosing the orientation of the dual edge $i$.

For a parabolic element, instead of formula (\ref{massshell2}), we have:
\beq p^2=0 \label{ll} \eeq
this corresponds to a massless particle. However, the mass of this particle is not
related to the deficit angle. On the other hand setting the parameter $m=0$ above does
{\em not} corresponds to a massless particle, since it characterizes the trivial holonomy. A massless particle has a genuine parabolic holonomy, its deficit angle is given as a function of the boost parameter:
\beq \cosh\eta\cos\frac{\alpha_{def}}{2}=\pm 1. \eeq
which is equivalent to the eq.(\ref{ll}). 
The remaining $N$ constraints are precisely the mass shell conditions and we assume that the Lorentz transformation $\Lambda_i$ corresponding to an edge incident to a one-vertex is never hyperbolic (which would be a tachyon with spacelike momentum). 

Switching to the scalar variables $L_i,\eta_i$ means that apart from the closure of the polygon $P$ all other above enumerated constraints are solved identically. We use them to determine the redundant angle variables which determine the directions of the edge vectors as well.  The closure of $P$ remains as a constraint, it can now be written as
\beq \sum_i L_i z_i(\eta_j)=0, \label{closure2} \eeq 
with the two vector $z_i=(E_{i+}+E_{i-})/L_i$ depending only on the boost parameters and  
\beq \sum_i \alpha_i(\eta_j)=(2E-2)\pi. \eeq
restricting the boost parameters. 
In a different form of the latter after plugging in $E=6g-3+2N$ one recognizes the Gauss-Bonnet theorem:
\beq H(\eta_j)\equiv\sum_{\cal V} \alpha_{def}=\sum_{\cal V}\left(\sum_{i\in {\cal V}}2\pi-\alpha_i(\eta_j)\right)=2\pi(2-2g). 
\label{hami} \eeq
where the summation ${\cal V}$ runs in the set of vertices and $i \in {\cal V}$ denotes the incidence of edge $i$ to the vertex ${\cal V}$.  
We have remaining symmetries corresponding to coordinate transformations of the chart:
\beq X\mapsto {\cal P}X=\Lambda X+a \eeq 
\beq {\cal P}_i\mapsto {\cal P}{\cal P}_i {\cal P}^{-1} \label{symmetry} \eeq
The translation part of the Poincare transformation ${\cal P}$ has trivial effect on the variables $\{\eta_i,L_i\}$ and so does a rotation $X\mapsto R(\psi)X$. So there are two independent nontrivial symmetries acting on the variables. One length parameter, say $L_1$, can be chosen freely, it corresponds to a constant reparametrization of the time coordinate \cite{evolution}. The number of physical degrees of freedom is the number of variables minus constraints minus symmetries:
\beq (2E-1)-3-2=12g-12+4N, \label{dof} \eeq
the dimension of the reduced phase space, twice the dimension of the moduli space of $N$ times punctured surfaces\footnote{apart from the exceptions enumerated in the beginning in section \ref{poligonmod}.} of genus $g$.

The dynamics associated to the chosen time slicing can be determined from the equations (\ref{wall}). The boost parameters are associated to the hyperplanes defined by these equations, so they are time independent. The edge vectors $E_{i\pm}$ move perpendicular to themselves with a constant velocity $\tanh \eta_i$. The variables are thus of action-angle type. We choose the convention that an edge $i_+$ (and its partner $i_-$) moves into the polygon if $\eta_i<0$. Within a finite amount of time if an edge shrinks to zero length or a concave angle hits an opposite side, nine different kinds of transitions can occur \cite{evolution}, the polygon may split into more polygons or a polygon may disappear, that is, the topology of the graph $\Gamma$ may change. Accordingly, the number of variables may change as well. The values of the new parameters are determined unambiguously by the geometry, the new consistency constraints and the mass shell constraints. We shall not deal with this issue further in this paper\footnote{In \cite{mi} it is proven for $g>1,N=0$ universes, that there is an invariant smooth hyperbolic surface ``triangulated" by $\gamma$ associated to each universe, which does not change during time translations and symmetries. In that case there is only one kind of transition which can occur, and that corresponds to changing the triangulation by deleting one diagonal of a quadrilateral and drawing the other one. The new boost parameters and lengths are determined by trigonometry of the quadrilateral. Similar effect is expected to happen in the more general case $N\neq 0$ at hand \cite{eninpr}.}.  
 
Let us choose the Hamiltonian to be the total deficit angle given by (\ref{hami}) as a function of the boost parameters.
This Hamiltonian generates the found time evolution \cite{canquant} if the symplectic structure is
\beq \{2\eta_i,L_j\}=\delta_{ij}. \label{sympstr} \eeq
\section{First order formalism in 2+1 gravity}
In the first order formalism of gravity the physical fields are the triad $e_{\mu}^a$ and the spin connection $\omega_{\mu}^a$. Locally both fields are one-forms with values in the Lie algebra of $so(2,1)\cong {\mathbb R}^{2,1}$. Greek indices are tangent space indices, latin indices $j,k...$ denote spatial components, latin indices from the beginning of the alphabet are indices from the Lie algebra, raised and lowered by the Minkowski metric with signature $(-,+,+)$. To avoid confusion the index $i$ will be used to index exclusively the finite number of variables corresponding to the edges of the graph $\Gamma$ of the previous section. The (Hilbert-Palatini) action reads (with $4\pi G=1$): 
\beq I=\frac{1}{2}\int  \epsilon^{\mu\nu\rho}\eta_{ab} e^a_{\mu}F^b_{\nu\rho}, \label{action} \eeq
where $F$ stands for the curvature of the spin connection:
\beq F^a_{\mu\nu}=\p_{\mu}\omega_{\nu}^a-\p_{\nu}\omega_{\mu}^a+\epsilon^{abc}\omega_{\mu b} \omega_{\nu c}. \eeq
The theory does not exclude degenerate triads, hence it is more general than the second order formalism in terms of the metric 
\beq g_{\mu\nu}=\eta_{ab}e^a_{\mu}e^b_{\nu}. \label{metric} \eeq
However in case of nondegenerate triads the same equations of motion lead to the same solutions as those of the usual Einstein-Hilbert action and the symmetry structure is also equivalent on shell. The physical interpretation for the triad is a local observer: in her Lorentz frame at the point $x$ the spacetime vector $v^{\mu}(x)$ is $e_{\mu}^a(x) v^{\mu}(x)$. With the help of the spin connection one can define the holonomy, that is the parallel transport operator along a curve $\alpha:[0,1]\rightarrow M^3$:
\beq U_{\alpha}(a,b)={\cal P}\exp\left(\int_0^1 ds \frac{d\alpha^{\mu}(s)}{ds}\omega_{\mu}^{a}T_a\right), \label{partrans} \eeq
where $\alpha(0)=a$, $\alpha(1)=b$, $(T_a)^b_{\;\,c}=\epsilon^b_{ac}$ and ${\cal P}$ stands for path ordering. If $v^d$ is a Lorentz vector over $b \in M^3$ then $U_{\alpha}(a,b)_{\;\;d}^c v^d$ is its parallel transported image over $a \in M^3$. 
 In order to work in the canonical formalism, one decomposes the action \`a la Arnowitt Deser and Misner and arrives at the following form:
\beq S=\int_I dt \int_{\Sigma} d^2x \eta_{ab}\epsilon^{0jk} (e_j^a \p_0 \omega_k^b+\omega_0^a D_j e_k^b+e_0^a F_{jk}^b) \label{admaction} \eeq
We assumed above the spacetime has the topology $M=\Sigma \times I$ where $\Sigma$ is a compact surface and $I$ is an interval.
The dynamical fields are the space components of the triad $e_i^a$ and the connection $\omega_i^b$, and one reads off from (\ref{admaction}) that they are canonically conjugate: 
\beq \{e_j^a(x),\omega_k^b(y)\}=\epsilon_{0jk}\eta^{ab}\delta(x,y), \eeq 
where $x$ and $y$ denote coordinates on the surface $\Sigma$. The space of classical solutions is spanned by the the solutions of the constraints:
\beq F_{jk}=0, \quad D_{[j}e_{k]}=0 \eeq
which are equations of motions coming from the variation with respect to the Lagrange multipliers $e_0^a$ and $\omega_0^a$, respectively. The torsion equations (Gauss constraint) generate the local $SO(2,1)$ transformations:
\beq \omega(=\omega^aT_a)\mapsto g^{-1}dg+g^{-1}\omega g, \quad e^a\mapsto (g^{-1})^a_{\;\,b} e^b \eeq
and the curvature equations generate local translations:
\beq \omega\mapsto\omega, \quad e^a_j\mapsto D_j\lambda^a. \label{lota} \eeq
For a pedagogical account on the topic, see e.g. \cite{matschullcanonical}. 
\section{Reduction to covariant variables and adding particles}
Our understanding of a point particle in 2+1 dimensions is based on the second order formalism. In \cite{djh} it has been shown, that with a stress energy of a massive point particle the solution to the Einstein equations is a cone. In Cartesian coordinates a particle is a point in Minkowski space with a wedge cut out and its edges glued together with time translation proportional to the spin of the particle and the angle of the wedge is proportional to the mass. To incorporate point particles in the theory the first step is to add the necessary terms to the action (\ref{admaction}), such that we obtain the correct equations of motion for them. The analysis has been done in both the mathematics and physics literature \cite{almal,soge,matschullonepart,matschullmp,erickarim}. Below we briefly summarize the procedure and the reduction of the action to a finite number of covariant variables. The references \cite{matschullonepart,matschullmp} are followed with some modifications in order to incorporate also higher genus surfaces as spacelike slices, that is, spaces with nontrivial tangent bundle. The last step, which is the main result of this paper will be done in section \ref{eredmeny}: arriving at the symplectic structure of the polygon model and its Hamiltonian as the constraint(s) remaining from the reduction. If there are point particles present, the first order and second order gravity are not even classically equivalent \cite{matschullehhp},  due to the different structure of large gauge transformations versus large diffeomorphisms. We do not deal with this fact here. We also restrict ourselves to the case of spinless particles.

Consider the equal time surface $\Sigma$ as a simply connected region $P=\Sigma/\Gamma$, with the graph $\Gamma$ of section \ref{poligonmod}, and pairwise identifications of the boundaries (which are the worldlines of the edges connecting vertices of $\Gamma$). On shell, the spin connection and the triad are given by pure gauge \cite{matschullonepart}:
\beq \omega^a_{\mu b}=(g^{-1}\partial_{\mu}g)^a_{\;\;b} \quad e_{\mu}^a=(g^{-1})^a_{\;\,b}\partial_{\mu}f^b \label{fields} \eeq 
with potentials $g:\mathbb{R}^2\times I\rightarrow SO(2,1)$ and $f:\mathbb{R}^2\times I\rightarrow {\mathbb R}^{2,1}$. They are defined on the universal cover, they are not single valued on the surface $\Sigma$. However, the fields should be single valued on $\Sigma$ which restricts the potentials. That is, most generally, on each pairs of edges $i_+$ and $i_-$ in $\partial P$ which correspond to the same edge $i$ of $\Gamma$ the following condition should hold:
\beq g_{i+}=\Lambda_{i} g_{i-}, \quad f_{i+}^a=\Lambda^{\,a}_{i\,b} (f_{i-}^b-a_{i}^b) \label{match} \eeq
with $\Lambda_i \in SO(2,1)$ and $a_i \in {\mathbb R^{2,1}}$. We need to impose the consistency condition: 
\beq \Lambda_i\Lambda_j\Lambda_k-id={\bf 0} \label{const1} \eeq
  for all triples $(i,j,k)$ corresponding to 3-vertices of $\Gamma$. This way the constraints in (\ref{admaction}) are solved everywhere, but at the 1-vertices. We want the 1-vertices to be point particles. To this end, we have to impose extra constraints: 
\beq tr\Lambda_j-2\cos m_j-1 \label{const2}=0 \eeq  
where $i$ runs in the set of labels of 1-vertices.
They make sure that the massive particles move along timelike geodesics, massless particles move along lightlike geodesics and there are conical singularities at their locations \cite{matschullonepart,matschullmp}. A second set of variables is given by the following formula: 
\beq E^a_{i\pm}\equiv\int_{i\pm}df^a_{i\pm}. \label{evec} \eeq  
The potential $f$ defines an embedding of $\Sigma$ to Minkowski space, and the ``edge vector" $E^a_{i\pm}$ is the relative position vector in the background Minkowski space of the two ends of edge $i_\pm$ of $\Gamma$. The image is a surface bounded by a (nonplanar) polygon, with pairwise identifications of the edges. The edge vectors are not independent, but satisfy
\beq E^a_{i+}=-\Lambda^{\,a}_{i\,b}E^b_{i-}, \label{holglue} \eeq
where the minus sign is due to the opposite orientation of the gluing. Finally there is a global constraint they also obey:
\beq C^a\equiv\sum_i \left(E^a_{i+}+E^a_{i-}\right)=0 \label{lorgen} \eeq
The computation of the symplectic potential is now identical to page 66 of \cite{matschullmp}.
Instead of repeating the calculation, we enumerate the notational differences. First, we have only one polygon, so $g_{\bigtriangleup}$ is just the potential $g$. Then $g_\lambda \leftrightarrow \Lambda_i$, Matschull labels the edges of $\Gamma$ with $\lambda$ and $z_\lambda\leftrightarrow E_{i-}$ with $\lambda>0$. Finally we use the vectorial representation of $SO(2,1)$, while he uses the spinorial. The result of the reduction reads:  
\beq \Theta=\int_{\Sigma}d^2x\epsilon^{0jk} \eta_{ab}\,d\omega^a_j e^b_k=-\sum_i \epsilon_{ac}^b(\Lambda_{i}^{-1}d\Lambda_{i})^a_{\;\,b}E^c_{i-}. \label{symppot} \eeq  
The above equation is the result of a page of calculation: Matschull plugs in the expressions (\ref{fields}), integrates over $P$, but the integrand is a total derivative. Hence, the two-dimensional integral becomes a sum over integrals for edges, which is again a total derivative and one finally obtains the rhs. of (\ref{symppot}). We will need the explicit expression for the Poisson bracket of the components of the global constraint (\ref{lorgen}). Since only the following brackets are nonzero among the edge vector components \cite{matschullmp}:
\beq \{E^a_{i\pm},E^b_{i\pm}\}=\epsilon^{ab}_c E^c_{i\pm}, \eeq
the global constraints (\ref{lorgen}) generate Lorentz transformations of the edge vectors and obey the same algebra:
\beq \{C^a,C^b\}=\epsilon^{ab}_c C^c. \eeq
The brackets of the rest of the constraints are analyzed in the literature \cite{matschullmp,thooftwael}, it will not be repeated here. Instead we go on with the symplectic reduction to recover the symplectic structure of the polygon model.
\section{'t Hooft polygon model in the first order formalism} \label{eredmeny}
Waelbroeck and Zapata have shown \cite{thooftwael} that a triangular lattice model based on the first order formalism also reduces to the polygon model if one introduces the scalar variables of the polygon model from the covariant parallel transport matrices and edge vectors, which solves some constraints identically. We will now show how the 't Hooft polygon model arises in the smooth first order formalism. First we require by means of a gauge transformation that the initial time surface is mapped to a {\em planar} polygon by the potential $f$ to Minkowski space. Then the scalar variables will be introduced by means of solving the constraints introduced in the previous section. One can proceed in two ways to recover the symplectic structure (\ref{sympstr}) of the scalar variables of the polygon model. The first is further reduction of the symplectic potential given by formula (\ref{symppot}). The second is the derivation of the Poisson brackets directly by appropriately expressing the scalar variables as functionals of the spin connection and the triad. Finally in the last section we identify the induced constraints, which provide dynamics for the model.   
\subsection{Phase space}
The edge vectors $E_{i\pm}^a$ and the holonomies $\Lambda_{i}$ are the finite number of covariant variables which contain the degrees of freedom of the system. To get to the polygon model we perform a local translation of (\ref{lota}) to obtain a planar polygon. 
Each edge vector $E_{i\pm}^a$ has to lie in a spacelike hyperplane in Minkowski space. Denote the unit timelike normal to this hyperplane by $n^a$. Let us define the boost parameters\footnote{Let us remark that for an edge $i$, it is possible to choose the gauge in such a way that $\Lambda_i$ is a pure boost hence tr$\Lambda_i=2\cosh 2\eta_i+1$ as proposed in \cite{richard}, but this cannot be done globally for all edges. It would mean that the ($\cosh$ of) the boost parameters are equal to traces of holonomies along closed curved, thus gauge invariants. We will see that this is not the case} and the lengths: 
\beq \cosh 2\eta_{i}=n_a\Lambda^{\,a}_{i\,b}n^b. \label{eta} \eeq
\beq L_{i}\equiv\sqrt{E^a_{i+} E^b_{i+} \eta_{ab}}=\sqrt{E^a_{i-} E^b_{i-} \eta_{ab}} \label{el} \eeq   
Now we can compute the symplectic structure departing from the rhs. of (\ref{symppot}). Let us choose a coordinate system in which the normal $n^a$ is purely timelike. The edge vectors are purely spacelike and we can use the explicit form given by (\ref{edgevec}) if $\Lambda_i$ is written as in eqn.(\ref{param}).  Plugging this into the term $\epsilon_{ac}^b(\Lambda_i^{-1}d\Lambda_i)^a_{\;\,b}E^c_{i-}$ the result of a direct calculation shows that the terms $d\phi$ and $d\phi'$ vanish and the coefficient of the $d(2\eta_i)$ term turns out to be $-L_i$.
Thus the symplectic potential is as expected:
\beq \Theta=-2\sum_i L_i d\eta_i \eeq
To understand better the origin of the result above, we present an alternative derivation of the Poisson brackets directly from the original fields of the theory. 
The Lorentz group element $\Lambda_i$ is the holonomy along the dual edge $\zeta_i$.  More precisely, if the points $a_-$ and $a_+$ are the starting and endpoints of a curve freely homotopic to the dual edge $\zeta$ with the same orientation, then the following relations are true: 
 \beq U_{\zeta\,b}^a(a_{-},a_{+})=g^{-1}(a_{-})g(a_{+})=g^{-1}(a_{-})\Lambda g(a_{-})=g^{-1}(a_{+})\Lambda g(a_{+}). \label{hollambda} \eeq
Note that the index $i$ is omitted from above to keep the notation simple. The holonomy can be also expressed in the usual way as the functional of the spin connection given by eqn.(\ref{partrans}). From now on we choose a fixed edge $i$ of $\Gamma$ and a dual edge of $\gamma$ denoted by $\zeta$.  The expression for the Poisson bracket of the associated length and boost parameter reads:
\beq \begin{array}{l}{\displaystyle
\{L,\cosh 2\eta\}=\{L,n_a(a_+) U_{\zeta\,b}^a(a_-,a_+) n^b(a_+)\}=} \\ {\displaystyle
\epsilon_{0jk}\eta^{cd} n_a(a_+) n^b(a_+)\int_{P}d^2x\frac{\delta L_i}{\delta e^c_j(x)}\frac{\delta U_{\zeta\,b}^a(a_-,a_+)}{\delta\omega_k^d(x)}} 
\end{array} \label{poisson} \eeq
where $n^a(a_+)=g^{-1}(a_+)^a_{\;\,b}n^b$.
In general, the functional derivative of the holonomy can be written as follows: %
\beq \frac{\delta U_{\zeta}}{\delta\omega_j^b(x)}=\int_{\zeta}ds \frac{dx^j(s)}{ds}\delta\left(x(s),x\right)U_{\zeta_1}T_b U_{\zeta_2} \label{hol} \eeq
where $\zeta_1$ is the segment of $\zeta$ until the point $x$, $\zeta_2$ is the segment of $\zeta$ from the point $x$ and $x(s)$ is the parametrization of the curve $\zeta$. The variation of the the length with respect to triad can be written using equations (\ref{fields}), (\ref{evec}), (\ref{el}) as
\beq \frac{\delta L}{\delta e^c_j(x)}=\frac{E_{b+}}{L}\int_i d\tau g(\tau)^b_{\;\,c} \frac{dx^j}{d\tau} \delta(x(\tau),x) \label{len}, \eeq
where $x(\tau)$ is a parametrization of the edge. 
Now we substitute (\ref{hol}) and (\ref{len}) to (\ref{poisson}) and integrate out one Dirac delta. If the integral for $\zeta$ is done first from the point $s_-=x(\tau)$ to its image $s_+$ along the curve homotopic to $\zeta$, then we have $\zeta_1=\emptyset$ and $\zeta_2=\zeta$, so we can write:
\beq \begin{array}{l}{\displaystyle \{L,\cosh 2\eta\}=} \\ 
{\displaystyle \int_i d\tau\int_{\zeta}ds\left[\epsilon_{0jk}\frac{dx^j(\tau)}{d\tau}\frac{dx^k(s)}{ds}\delta\left(x(\tau),x(s)\right)\right]\frac{E^b_+ g(s_+)^{\;\,a}_b}{L} n_c(s_+)\left(T_a U(s_-,s_+)\right)_{\;\;b}^c n^b(s_+)}. \label{bracket} \end{array} \eeq
The expression outside the square bracket is independent of $s$, since we can rewrite it using the notation $(v\times w)_a=\epsilon_{abc} v^b w^c$ as 
\beq g^{-1}(s_+)E_+)^a (g^{-1}(s_+)n \times g^{-1}(s_+)\Lambda n)_a=E^a_+(n\times\Lambda n)_a,\eeq
 because both the vector product and the Minkowski scalar product are invariant under Lorentz transformations. The integral of the square bracket in equation (\ref{bracket}) is the homotopy invariant oriented intersection number of the edge $i$ and the closed curve $\zeta$. It is one for the pairs $(i,\zeta_i)$ and zero otherwise. The last ingredient is to verify the following formula:
\beq n\times\Lambda_i n=\sinh(2\eta_i) E_{i+}. \eeq
$E_{i+}$ is clearly orthogonal to both $n$ and $\Lambda_i n$ due to (\ref{holglue}) and the scalar factor in the previous equation is also not difficult to check using the definition (\ref{eta}). We have thus arrived at the desired result:
\beq \{L_i,\cosh 2\eta_j\}=\delta_{ij}\sinh 2\eta_i \rightarrow \{L_i,2\eta_j\}=\delta_{ij} \eeq
\subsection{Dynamics}
At the intermediate stage of the symplectic reduction, the finite number of covariant variables $\Lambda_i$ and $E_i$ have to satisfy a number of constraints, given by (\ref{const1}), (\ref{const2}), (\ref{holglue}) and (\ref{lorgen}). Changing to the scalar variables $L_i,\eta_i$ identically solves the first three, if the triangle inequalities
\beq \vert\eta_i\vert+\vert\eta_j\vert\geq\vert\eta_k\vert \eeq
are satisfied, for every vertex ${\cal V}=\{i,j,k\}$, and for every permutation of $\{i,j,k\}$. The explanation is the following: Eqn.(\ref{const1}) and (\ref{const2})  determine the angles of the polygon in terms of $\eta_i$, eqn.(\ref{holglue}) determines the direction of the edges. In reference \cite{thooftwael} it is argued, that the Gauss constraint is solved by introducing the scalar variables and there remain induced curvature constraints, but the situation is on the contrary: the Gauss law and only that remains after the reduction. Namely, the closure of the polygon $P$: (\ref{lorgen}) written as (\ref{closure2}) with only two independent components and the sum of angles in the Euclidean polygon equivalent to the Gauss-Bonnet theorem for the surface it represents, given by eqn.(\ref{hami}). 
The latter is the Hamiltonian, which generates a time evolution that preserves the closure of the polygon by construction \cite{canquant}, thus
\beq \{H,C^a\}=\sum_i \dot{L}_i z_i=0, \eeq
where the dot now indicates the time evolution induced by $H$.
The fact that $\{C^a,C^b\}=\epsilon_c^{ab} C^c$ does not get modified is difficult to check explicitly. However, the correct number of degrees of freedom comes out as in eqn.(\ref{dof}) only if all three independent constraints $H$, $C_1$ and $C_2$ are first class and generate time translation and two independent Lorentz transformations respectively (since a rotation is factored out by using lengths and angles instead of edge vectors). In other words the constraint algebra closes and the most general time evolution is generated by the following Hamiltonian:
\beq H'=a\,H+b\,C^1+c\,C^2. \eeq
\section{Discussion}
In this paper, the 't Hooft model of globally hyperbolic 2+1 dimensional Lorentzian universes with zero cosmological constant and compact spacelike part has been derived in terms of the fields of the first order formalism of gravity. We could reduce the originally infinite dimensional phase space spanned by the spin connection and the triad to the space of finite number of scalar variables in a simple way for the generic case of arbitrary genus and number of point particles. The induced symplectic structure has been derived both directly from the smooth fields and from a finite set of covariant variables of Matschull, Waelbroeck and Zapata without the need of introducing any lattice. Apart from showing the connection of different approaches to 2+1 gravity, interpreting the polygon model directly as a gauge theory has another advantage. While the original treatment in terms of geometric structures has a ``rigid" time evolution generated by $H$ and induced by the slicing of space time, in gauge theory, there is an arbitrariness of taking any linear combination of the constraints to be the Hamiltonian. One may use this freedom to prove that one-polygon tessellation is generic and transitions corresponding to splitting (and disappearance) of polygons may be completely avoided. Another aspect which makes it worths to pursue progress in this direction is the cotangent bundle structure of the polygon model. It is a convenient starting point for a quantum theory. However this has only been shown \cite{mess} and an explicit section given \cite{mi} for the case of no point particles. For the case of point particles there is no complete classification of the phase space \cite{benguad}.        
\section*{Acknowledgements}
Thanks to R. Loll, D. N\'ogr\'adi and  B. Schroers for valuable discussions and to the organizers of the workshop ''Towards the quantum geometry of hyperbolic 3-manifolds" in Potsdam for their hospitality.   

\begin{thebibliography}{100}
\bibitem{causality} 't Hooft, G., Causality in (2+1)-dimensional 
gravity, Class.\ Quant.\ Grav.\ {\bf 9} (1992) 1335-1348.
\bibitem{evolution} 't Hooft, G., The evolution of gravitating point
particles in 2+1 dimensions, Class.\ Quant.\ Grav.\ {\bf 10} (1993) 1023-1038.
\bibitem{gth} 't Hooft, G., Classical N-particle cosmology in
2+1 dimensions,  in 't Hooft, G. (ed.): Under the spell of the gauge
principle, 606-618, and Class.\ Quant.\ Grav.\ {\bf 10} (1993) Suppl.,
79-91. 
\bibitem{canquant} 't Hooft, G., Canonical quantization of gravitating
  point particles in 2+1 dimensions, Class.\ Quant.\ Grav.\ {\bf 10}
  (1993) 1653-1664 [arXiv: gr-qc/9305008].
\bibitem{achtow} Achucarro, A., Townsend, P., Chern-Simons Action For Three-Dimensional Anti-De Sitter Supergravity Theories, Phys.\ Lett.\ B {\bf 180} (1986) 89.
\bibitem{witt} Witten, E.,
(2+1)-dimensional gravity as an exactly soluble system,
Nucl.\ Phys.\ B {\bf 311} (1988) 46-78.
\bibitem{matschullehhp} Matschull, H.-J., On the relation between 2+1 Einstein gravity and 
Chern Simons theory, Class.Quant.Grav.16:2599-2609,1999, 
[arXiv:gr-qc/9903040].
\bibitem{soge} Sousa Gerbert, P.,de, On Spin and (Quantum) Gravity in 2+1 Dimensions, 
Nucl.Phys.B346:440-472,1990, 
\bibitem{carlip} Carlip, S., Quantum Gravity in 2+1 Dimensions: The Case of a Closed Universe, gr-qc/0409039 
\bibitem{menotti}
Cantini, L., Menotti, P., Seminara, D.,
Hamiltonian structure and quantization of 2+1 dimensional gravity 
coupled to particles,
Class.\ Quant.\ Grav.\  {\bf 18} (2001) 2253-2276
[arXiv: hep-th/0011070].
\bibitem{bernd} Schroers, B., Meusberger, C., Poisson structure and symmetry in the Chern-Simons formulation of (2+1)-dimensional gravity, Class.\ Quant.\ Grav.\  {\bf 20} (2003) 2193
[arXiv:gr-qc/0301108].
\bibitem{waelmod} Waelbroeck, H., 2+1 lattice gravity,
  Class.\ Quant.\ Grav.\ {\bf 7} (1990) 751-769.
\bibitem{unne} Unruh, W. G., Newbury, P., Solution to 2+1 gravity in the dreibein formalism, Phys.Rev.D.48,2686-2701,1993 [arXiv:gr-qc/9307029].
\bibitem{matschullonepart} Matschull, H.-J., Quantum Mechanics of a Point Particle in 2+1 Dimensional Gravity, Class.Quant.Grav.15:2981-3030,1998, [arXiv:gr-qc/9708054].
\bibitem{matschullmp} Matschull, H.-J., The Phase Space Structure of Multi Particle Models in 2+1 Gravity, Class.Quant.Grav.18:3497-3560,2001, [arXiv:gr-qc/0103084].
\bibitem{erickarim} Buffenoir.E, Noui, K., Unfashionable Observations About Three Dimensional Gravity, arXiv:gr-qc/0305079.
\bibitem{exceptions} Nelson, J.E., Regge, T., 
Phys.\ Rev.\ D {\bf 50} (1994) 5125-5129
[arXiv: gr-qc/9311029],
Ashtekar, A., Loll, R.,
Class.\ Quant.\ Grav.\ {\bf 11} (1994) 2417-243,
[arXiv: gr-qc/9405031],
Moncrief, V., J.\ Math.\ Phys.\  {\bf 30} (1989) 2907-2914,
Hosoya, A., Nakao, K.-I.,
Class.\ Quant.\ Grav.\  {\bf 7} (1990) 163-176.
\bibitem{witttc} Witten, E., Topology Changing Amplitudes In (2+1)-Dimensional Gravity, 
Nucl.\ Phys.\ B {\bf 323} (1989) 113.
\bibitem{freidel} Freidel, L., Louapre, D., Ponzano-Regge model revisited. I: Gauge fixing, observables and interacting spinning particles, 
arXiv:hep-th/0401076.
\bibitem{karimperezpart} Noui, K., Perez, A., Three dimensional loop quantum gravity: Coupling to point particles,
arXiv:gr-qc/0402111.
\bibitem{thooftwael}
Waelbroeck, H., Zapata, J.A., 2+1 covariant lattice
theory and 't Hooft's formulation, Class.\ Quant.\ Grav.\ {\bf 13} (1996)
1761-1768 [arXiv: gr-qc/9601011].
\bibitem{djh} Deser, S., Jackiw, R., 't Hooft, G., Three-Dimensional Einstein Gravity: Dynamics of Flat Space, Ann. Phys. {\bf 152}, 220-235 (1984)
\bibitem{matschullcanonical} Matschull, H.-J., Three Dimensional Canonical Quantum Gravity, Class.Quant.Grav.12:2621-2704,1995
[arXiv:gr-qc/9506069].
\bibitem{barbot} Barbot, T., in preparation
\bibitem{mi} Kadar, Z., Loll, R. (2+1) gravity for higher genus in the polygon model, Class.Quant.Grav. 21 (2004) 2465-2491
[arXiv:gr-qc/0312043].
\bibitem{almal} Alekseev, A. Yu., Malkin, A. Z., Symplectic structure of the moduli space of flat connection on a Riemann surface, Commun.Math.Phys. 169 (1995) 99, 
[arXiv:hep-th/9312004].
\bibitem{eninpr} Kadar, Z., in preparation
\bibitem{richard} Livine, E.-R., Loop quantum gravity and spin foam: covariant
methods for the nonperturbative quantization of general relativity,
PhD thesis [arXiv: gr-qc/0309028].
\bibitem{mess} Mess, G., (1990). Lorentz spacetimes of constant curvature, Institute des Hautes Etudes Scientifiques preprint IHES/M/90/28
\bibitem{benguad}
Benedetti, R., Guadagnini, E., Geometric Cone Surfaces and (2+1) - Gravity coupled to Particles, Nucl.Phys.B588:436-450,2000, 
[arXiv:gr-qc/0004041].
\end{thebibliography}
\end{document}